\documentclass[twocolumn,english,aps,prb,superscriptaddress,bibnotes,amsmath,amssymb,floatfix,nobalancelastpage]{revtex4-2}

\usepackage{graphicx}
\usepackage{dcolumn}
\usepackage{bm}
\newcommand{\Fref}[1]{Fig.~\ref{#1}}
\usepackage[colorlinks=true,citecolor=blue,linkcolor=magenta]{hyperref}

\begin{document}

\preprint{APS/123-QED}

\title{Photonic Generation of Radar Signals with 30 GHz Bandwidth\\ and Ultra-High Time-Frequency Linearity}
\author{Ziqian Zhang}
\author{Yang Liu}
\email{yang.liu@sydney.edu.au}
\author{Benjamin J. Eggleton}
\affiliation{Institute of Photonics and Optical Science (IPOS), School of Physics, The University of Sydney, NSW 2006, Australia}
\affiliation{The University of Sydney Nano Institute (Sydney Nano), The University of Sydney, NSW 2006, Australia}

\date{\today}

\begin{abstract}
Photonic generation of radio-frequency signals has shown significant advantages over the electronic counterparts, allowing the high precision generation of radio-frequency carriers up to the terahertz-wave region with flexible bandwidth for radar applications. Great progress has been made in photonics-based radio-frequency waveform generation. However, the approaches that rely on sophisticated benchtop digital microwave components, such as synthesizers and digital-to-analog converters have limited achievable bandwidth and thus resolution for radar detections. Methods based on voltage-controlled analog oscillators exhibit high time-frequency non-linearity, causing degraded sensing precision. Here, we demonstrate, for the first time, a photonic stepped-frequency (SF) waveform generation scheme enabled by MHz electronics with a tunable bandwidth exceeding 30 GHz and intrinsic time-frequency linearity. The ultra-wideband radio-frequency signal generation is enabled by using a polarization-stabilized optical cavity to suppress intra-cavity polarization-dependent instability; meanwhile, the signal's high-linearity is achieved via consecutive MHz acousto-optic frequency-shifting modulation without the necessity of using electro-optic modulators that have bias-drifting issues. We systematically evaluate the system's signal quality and imaging performance in comparison with conventional photonic radar schemes that use high-speed digital electronics, confirming its feasibility and excellent performance for high-resolution radar applications.   
\end{abstract}

\maketitle


\textit{Introduction}.\textbf{---}Radar sensing has been progressing into the millimeter-wave (MMW, 30-300 GHz) and terahertz-wave (THz-wave, 0.1-10 THz) regions to operate with ultrawide bandwidth for growing demands of high spatial resolution imaging in real-world applications, such as non-destructive testing, automotive driving assistance, industrial quality inspection, and non-invasive medical imaging \cite{Duling2009,Gowen2012,Caruso2015}. However, the development of wideband radar has posed significant challenges to conventional electronic technologies, especially in the synthesis of ultra-broadband signals. In particular, the bandwidth of direct digital synthesizers and digital-to-analog converters (two commonly used and essential components for signal generation) are constrained to a few gigahertz by the limited clock speed \cite{Richards2010}. Moreover, these devices have shown significant degradation in both efficiency and noise level when approaching higher frequencies via multi-stage frequency up-conversion \cite{Tonda-Goldstein2006}. Additionally, multi-stage frequency multiplexing and spectrum stitching for bandwidth broadening introduce noise and spectrum spur induced by device nonlinearity and interference \cite{Serafino2020a}, which compromises overall sensing accuracy and performance.

Photonics-assisted radar shows the potentials to overcome the drawbacks in the electronic counterparts mentioned above, especially in bandwidth broadening \cite{Pan2020}, low-noise up-conversion \cite{Kittlaus2021}, and simple down-conversion and demodulation \cite{Serafino2019}. However, existing photonic approaches for radar signal generation still rely on bulky, high-speed benchtop electronics, or lossy and elaborately biased electro-optic modulators (EOMs), limiting the achievable bandwidth, long-term operating stability, and ultimately the practicality \cite{Fu2013,Pan2020}. Alternative approaches using dispersion-based time-stretch \cite{Zhang2014} and frequency-sweeping light sources \cite{Zhou2016,Hao2018} have also shown promising bandwidth capacity. However, these techniques remain challenging to meet simultaneously the high frequency-time linearity for accurate ranging without using pre-distorted control signals for linearity compensation and the wide bandwidth for a fine spatial resolution to cope with the real-world, resolution-demanding applications. 

In this paper, we demonstrate, for the first time, an MHz electronics-enabled photonic synthesizing of stepped-frequency (SF) signal with tunable bandwidth exceeding 30 GHz and an inherent high frequency-time linearity. Because of using a polarization-maintaining optical cavity against the ambient environment perturbation and the stable acousto-optic frequency-shifting modulation without the bias-drifting issues from EOMs, the demonstrated system reaches a signal-to-noise ratio (SNR) of above 34 dB in signal generation comparable to those generated by high-end benchtop electronics. Simultaneously, the constant frequency shift and cavity round-trip time ensure an inherent high frequency-time linearity with a maximum deviation below 2.5 MHz throughout the bandwidth for accurate detection. We experimentally compare a radar system using the demonstrated scheme with a conventional photonic radar that relies on a high-speed waveform generator, clearly showing its comparable performance in radar imaging. With the achieved bandwidth exceeding 30 GHz, this demonstrated approach provides a viable basis for ultra-wideband microwave waveform synthesis for future ultra-high-resolution MMW and THz-wave radar systems.

\begin{figure}[t!]
\centering
\includegraphics[width=\linewidth]{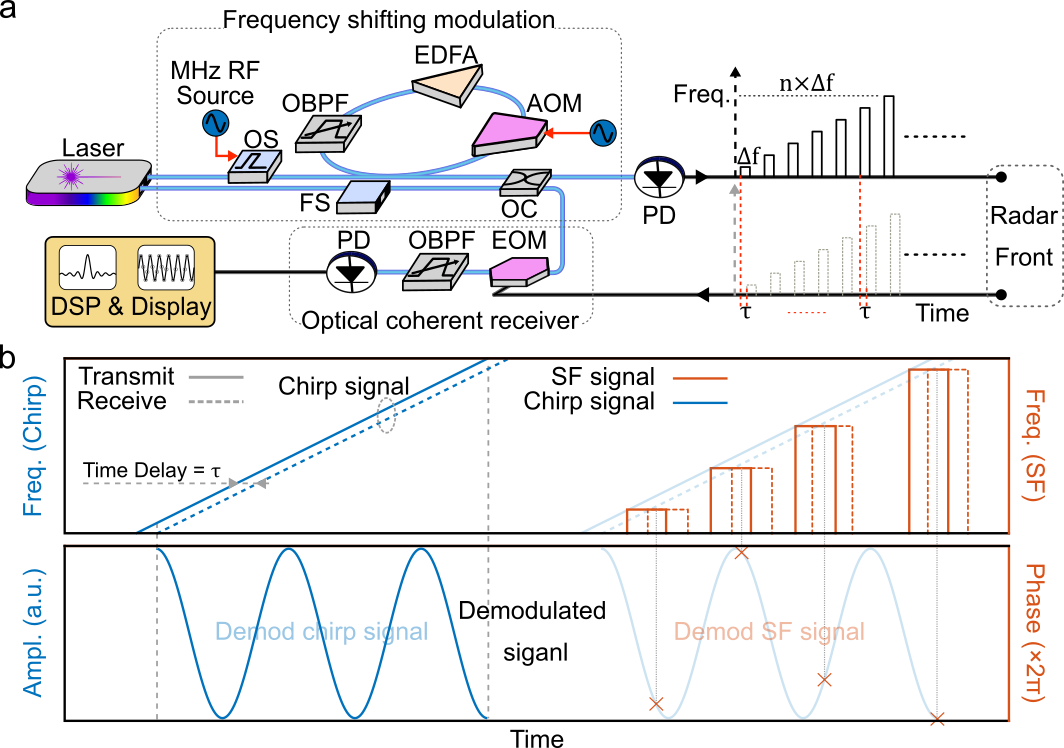}
\caption{(a) Schematic of the demonstrated SF radar using optical frequency-shifting modulation. (b) Principle comparisons of radar systems using linear-frequency modulated (LFM or chirp) and SF signals. RF, radio-frequency; OS, optical switch; OBPF, optical bandpass filter; EDFA, erbium-doped fiber amplifier; AOM, acousto-optic modulator; FS, optical frequency shifter; OC, optical coupler; PD, photodetector; DSP, digital signal processing; SF, stepped-frequency; $\tau$, the time delay of the reflected detecting signal.}
\label{fig1}
\end{figure}

\textit{Experiments and Results}.\textbf{---}The demonstrated SF waveforms are generated through recirculating an optical rectangular pulsed signal originated from chopping a continuous wave (CW) laser in a frequency-shifting loop (FSL) \cite{Zhang2020a}, diagrammed in \Fref{fig1}(a). An acousto-optic modulator (AOM) precisely shifts the pulse frequency for each round-trip, offering inherent ultra-high frequency-time linearity and ultra-fast frequency shifting. Moreover, the AOM shifted signal avoids the generation of spectrum harmonic spurs, which is an outstanding advantage over the approach using EOMs for single-sideband modulation (SSB) that suffers from bias-drifting and parasitic harmonics \cite{Fu2013,Zhang2021}. An optical bandpass filter (OBPF) is used to determine and tune the bandwidth of the optical and thus the RF SF signals to satisfy diverse resolution requirements (range resolution is defined as $c/2B$, where $B$ is the synthesized bandwidth of the signal). Meanwhile, an erbium-doped fiber amplifier (EDFA) compensates for the optical modulation and propagation energy losses. It is worth mentioning that the pulse dwelling time set by an optical switch (OS) should be less than the round-trip time of the FSL to avoid inter-pulse cross-talk. A simple optical-RF up-conversion is realized by mixing the optical SF signal with a frequency-shifted CW laser signal in a photodetector (PD). Finally, a coherent optical receiver will demodulate the received signal for ranging and imaging \cite{Kikuchi2016}.  

\Fref{fig1}(b) shows the principle of the SF signal, in comparison to linear frequency modulated (LFM) signals, which are both widely used as pulse-compression waveforms for radar sensing owing to their benefits of higher mean powers while sustaining superior resolution \cite{Ozdemir2012}. The operation principles of the SF signal and the LFM signal have fundamental but subtle differences; intuitively, the SF signal is a discretely sampled version of the LFM signal, as shown in \Fref{fig1}(b). In radar applications, the SF radar encodes the time delay ($\Delta \tau$) of the reflected signal as phase differences with respect to the reference signals ($\omega_{SF} \propto 2\pi n \Delta f \Delta \tau$, where $\Delta f$ is the incremental frequency, and $n = 1,..., N, N$ is the total number of frequency steps). The LFM radar forms a new oscillating frequency proportional to the round-trip time $\Delta \tau$ ($\omega_{LFM} \propto 2\pi k \Delta \tau t$, where $k$ is the chirp rate \cite{Levanon2002}), as depicted in \Fref{fig1}(b). Advantageously, SF radars require a much lower digitizing speed (down to only one sample per step-time) than the LFM radar receivers bounded by the Nyquist sampling theorem. This characteristic allows the photonic SF radars to generate much less data while sustaining the same resolution, enabling low-consumption edge computing for resolution-demanding applications such as hand gesture recognition and object identification \cite{Jia2021a}. Moreover, the SF waveforms have shown other advantages over the LFM signals. For instance, it possesses an increased dynamic range due to each frequency step's narrow instantaneous noise bandwidth for signal processing while preserving the range resolution and the overall bandwidth \cite{Caruso2015}.

\begin{figure}[t!]
\centering\includegraphics[width=\linewidth]{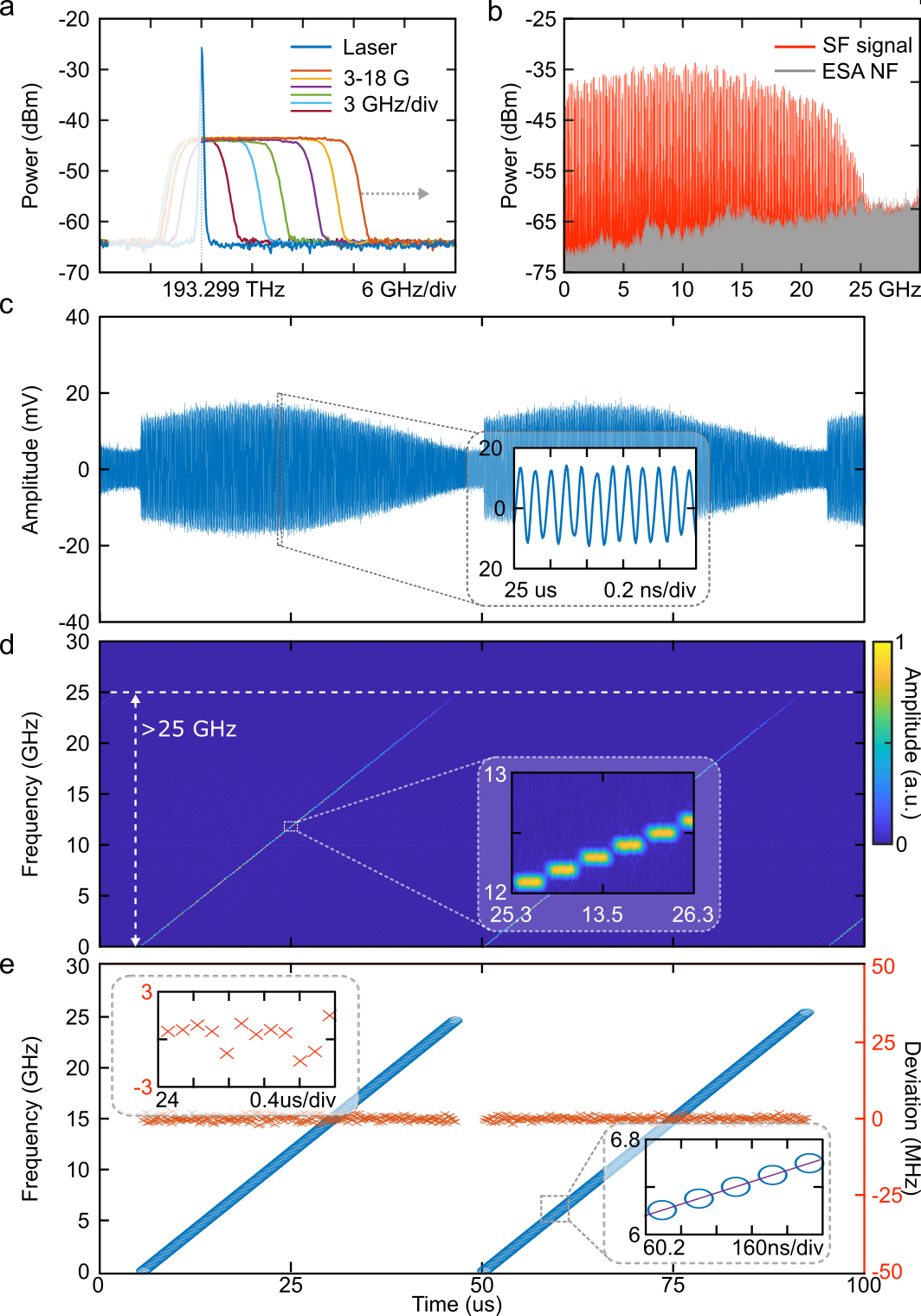}
\caption{(a) Optical spectra showing the laser frequency and several passbands of the OBPF from 3 to 18 GHz. (b) The baseband, RF-domain spectrum of the SF signal with the measuring apparatus's noise floor (NF). (c) A time-domain plot of the SF signals with 250 steps, 100 MHz frequency shift, and a total synthesized bandwidth of 25 GHz. (d) The time-frequency plot of the 25 GHz bandwidth SF signal using a 25 ns sliding time window. (e) SF linearity analysis based on extracting the peak values of the time-frequency data with one sample each frequency step and a constant sampling time interval. A first-order polynomial fitting is applied to those data for calculating the deviation.}
\label{fig2}
\end{figure}

\Fref{fig2} shows a typical waveform and spectrum of the generated SFCW signal to investigate the quality and linearity. One tap of the CW laser mixes the optical SF signal in a PD to generate the SF signal in the RF baseband. \Fref{fig2}(b) plots the spectrum of the SF signal in the RF-domain as well as the noise floor (input disconnected) of the measuring apparatus (Agilent E4448A). It confirms the 25 GHz SF bandwidth, tantamount to a theoretical range resolution of 6 mm. The temporal waves and the corresponding frequency-time relations are presented in \Fref{fig2}(c) and (d), respectively. These plots demonstrate that each frequency step only contains a single oscillating frequency and is progressively shifted after a specific dwell time (i.e., the round-trip time of the FSL). The time-domain envelope roll-off at the higher frequency side is mainly contributed by the roll-off of the OBPF, which is confirmed by the spectral measurements in both the optical (\Fref{fig2}(a)) and RF domain (\Fref{fig2}(b)). A further linearity analysis is carried out by comparing the difference between the experimental results (extracting the frequency of the time-domain signal in each frequency step) and the corresponding values of the first-order polynomial fitting, which is shown as the deviation in \Fref{fig2}(e). The linearity analysis reveals a ladder-like frequency-time feature with a maximum deviation below 2.5 MHz throughout the 25 GHz bandwidth. The deviation is overestimated and is limited by the resolution of the digital processing within a finite time window. This feature ensures the high linearity of the SF format, which is hard to achieve using laser sweeping or voltage-controlled oscillators without an active feedback control loop or pre-distorted RF signals generated from benchtop electronic units.

\begin{figure}[t!]
\centering\includegraphics[width=\linewidth]{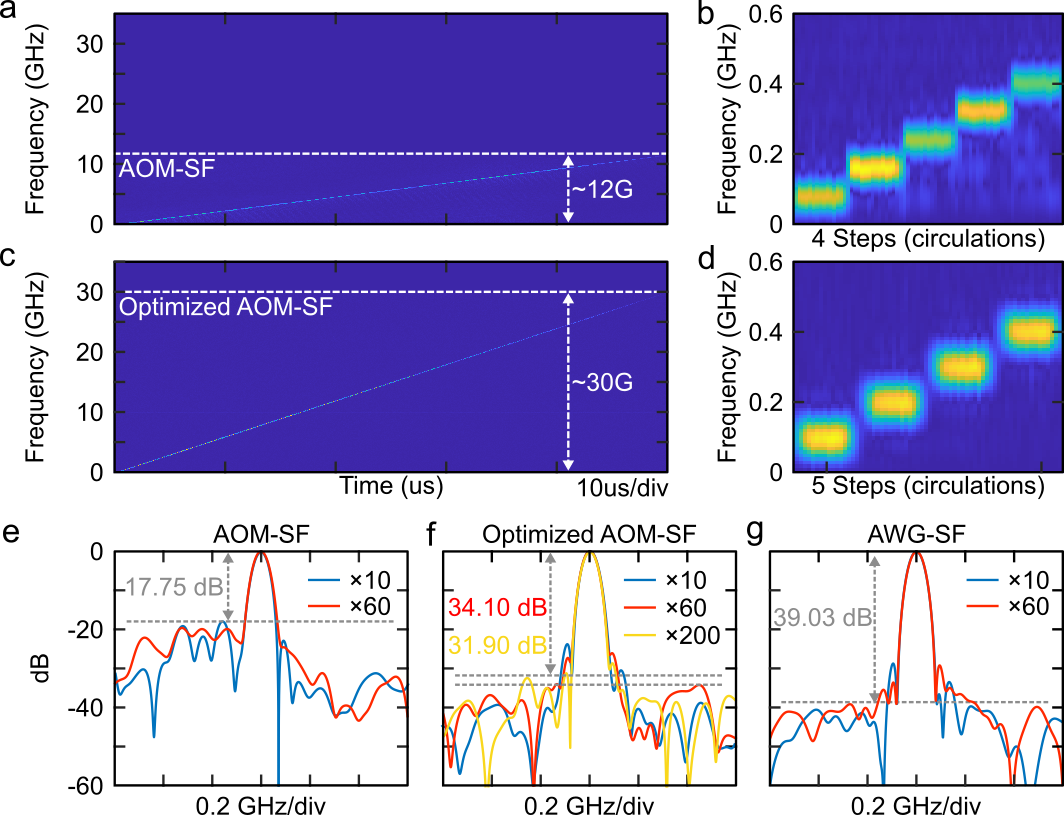}
\caption{Bandwidth and signal-to-noise ratio (SNR) comparison between the systems using single-mode fibers (SMFs) and polarization-maintaining fibers (PMFs). (a) Time-frequency plot of the signal based on un-optimized FSL using SMFs with a bandwidth of 12 GHz. (b) An insight of the 12 GHz signal. (c) Time-frequency plot of the signal based on an optimized FSL using PMFs with a bandwidth exceeding 30 GHz. (d) An insight of the 30 GHz signal. (e) SNR plots of the SMF-based system at two frequency instances, i.e., 10-time recirculation (blue) and 60-time recirculation (red) with SNRs around 17.75 dB. (f) SNR plots of the PMF-based system at the same frequency instances showing an SNR of 34.10 dB at 60-time recirculation. 200-time recirculation is plotted (yellow) with an SNR of 31.90 dB. (g) SNRs of the SF signal generated using an AWG at the same frequency instances.}
\label{fig3}
\end{figure}

Bandwidth and frequency shift tunability (changing $\Delta f$) are prominent metrics when deploying SF signals for various applications as two key sensing parameters are decided accordingly, i.e., the range resolution, $c/2B$ and ambiguity, $c/(2 \times \Delta f)$. In principle, the demonstrated system can achieve arbitrary bandwidth tuning by changing the passband and central frequency of the OBPF, thereby enabling range resolutions down to mm level. However, in practice, such broadband synthesizing is challenging for single-mode fibers (SMFs), especially when polarization will be further scrambled in the optical cavity, deteriorating the phase stability, signal coherence, and ultimately the sensing performance. As shown in \Fref{fig3}(a), the time-frequency plot of the SF signal generated through an un-optimized FSL using SMF is challenging to achieve a bandwidth of 12 GHz. \Fref{fig3}(b) provides an insight into the time-frequency plot, revealing the amplitude fluctuations across different frequency steps, caused by the polarization and gain instability.

In contrast, the signal bandwidth and stability can be significantly increased using polarization-maintaining fibers (PMFs) and components that can minimize the polarization scrambling, and thus help achieve the ultra broad bandwidth synthesizing over 30 GHz. \Fref{fig3}(c) shows an $>$30 GHz bandwidth signal synthesizing using PMFs against polarization scrambling caused by polarization-dependent reflections originating from optical couplers and the ambient environment, such as temperature and vibration. Compared to \Fref{fig3}(b), the optimized system also reduces the background noise by using an OS with a high on-off extinction to suppress the seed pulse tails recirculating in the FSL. As a result, \Fref{fig3}(d) proved that the improved system performs a more stable pulse-circulation with much less interference from inter-pulse overlapping and noise accumulation.

To provide more insights into the signal quality, we compare the SNR of the signal generated using different schemes. Fourier analysis using 25 ns, time-domain signal clips between SMF-based system and PMF-based system are shown in \Fref{fig3}(e) and \Fref{fig3}(f), respectively. The results are sampled at two instances, i.e., 10-time (blue) and 60-time (red) recirculation, revealing a more than 14 dB SNR improvement with an SNR of 34. In the meantime, the result of 200-time recirculation in the optimized system exhibited in \Fref{fig3}(f) (yellow) also proves that the SNR does not obviously degrade as the pulse recirculation time increases, indicating a low stability penalty from the FSL noise accumulations. Moreover, we compared the SF signal generated between the PM-based FSL and an arbitrary waveform generator (AWG, Keysight M8195A 65GSa/s). Signal quality analysis in \Fref{fig3}(g) shows that the demonstrated system sustains an SNR above 34 dB after 60-time recirculation in the PMF-based FSL compared with 39 dB SNR from the high-end electronics. The results also proved that the demonstrated system could replace high-speed, noisy electronic synthesizers to generate ultra-broadband signals directly at the carrier band for accurate MMW and THz-wave sensing. 

\begin{figure}[t!]
\centering\includegraphics[width=\linewidth]{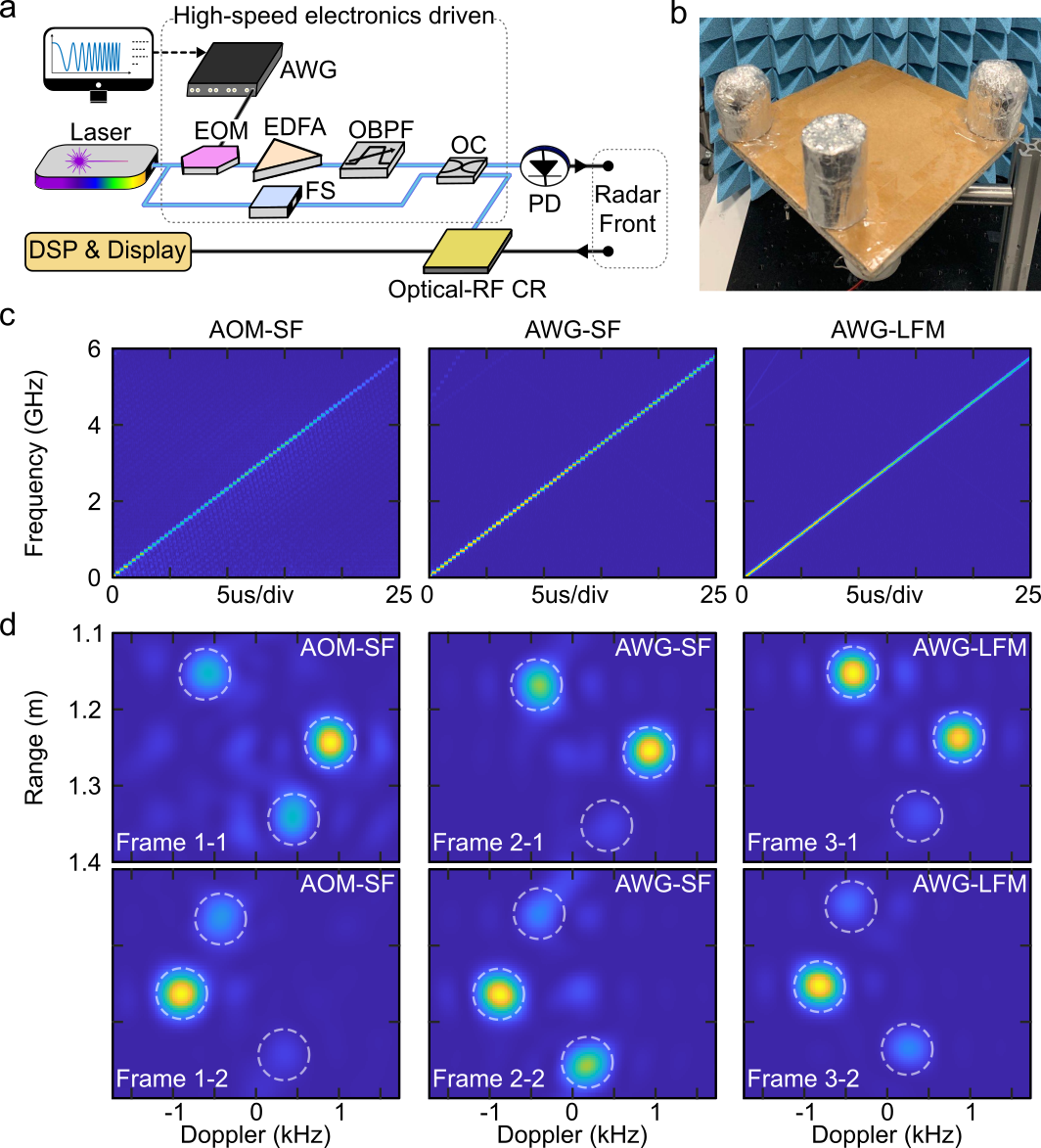}
\caption{(a) Schematic of a photonic radar system driven by a high-speed signal generator, which is used to compare the imaging performance with the demonstrated radar system. (b) The target for radar imaging with three-cylinder objects mounted on a rotating platform. (c) The signals used for the imaging performance comparisons. (d) 2D imaging results. EOM, electro-optic modulator; CR, coherent receiver.}
\label{fig4}
\end{figure}

In order to demonstrate the competitive performance and feasibility, we compare radar imaging performance based on the SF signal generators with those using high-speed electronic AWGs as shown in \Fref{fig4}. 2D imaging experiments are conducted based on the inverse synthetic aperture radar (ISAR) technique. Comparison experiments that use ultra-fast electronics (Keysight M8195A, 65GSa/s, $\sim$25 GHz analog bandwidth) for generating both SF and LFM signals are carried out based on the schematic shown in \Fref{fig4}(a) while keeping the same bandwidth (5.76 GHz), step numbers (72 steps), repetition rate (25 us), and RF radiation power ($\sim$ 10 dBm), as illustrated in \Fref{fig4}(c). It should be noted that the signal bandwidth in the demonstrations were chosen due to the bandwidth availability of the RF antenna, without losing comparison generality. Wideband-RF signals generated by the AWG modulate a CW laser carrier through a single-sideband modulation before beating with a separate optical carrier for optical-RF up-conversion. As shown in \Fref{fig4}(b), three-cylinder objects mounted on a rotating platform are used as the target. ISAR imaging results in \Fref{fig4}(d) are chosen at two particular instances for an adequate performance illustration and comparison, proving that the demonstrated system is reliable for imaging moving objects. Notably, the demonstrated system successfully reconstructed ISAR images of the objects without distinguishable differences from those based on the AWG. It should be noted that a minor frequency step ($\Delta f$) can be achieved by cascading two AOMs with the opposite frequency shift (e.g., a frequency shift from -10 to +10 MHz has been demonstrated in \cite{Billault2021}) to achieve an unambiguous range over 100 meters, which is comparable to the 120 meters unambiguous range window (1.25 MHz frequency shift) from the CARABAS system - a very early airborne synthetic aperture radar that employed SF signals - mounted on an aircraft \cite{Gustavsson1990}. 

\textit{Conclusions}.\textbf{---}In conclusion, we demonstrated a photonic SF waveform generation with a tunable bandwidth of $>$30 GHz and intrinsic high-linearity for high-resolution and accurate radar detections. By stabilizing polarization in the optical FSL loop, we significantly improve the SNR of the SF signals from 17 dB to 34 dB with a low noise accumulation penalty, allowing imaging performance comparable to those using high-speed apparatus. The demonstrated system enables high-resolution radar sensing and represents an attractive combination of wideband signal synthesizing, high SNR, and reduced hardware requirements. This work serves as a pilot study and experimental basis for further chip-based radar integration using on-chip devices \cite{Shao2020a}, opening the door to future ultra-high-resolution, miniaturized, and mobile millimeter-wave devices with prime performance and flexibility.

\bigskip

\noindent\textbf{Funding:} This work was funded by Australian Research Council Discovery Project (DP200101893).

\noindent\textbf{Data Availability Statement:} The data that support the findings of this study are available from the corresponding author upon reasonable request.

\noindent\textbf{Author Contributions:} Y.L. and B.J.E. conceived the project; Z.Z. and Y.L. designed the system structure; Z.Z. conducted the experiments; Z.Z. and Y.L. performed data analysis. Z.Z. wrote the manuscript with contribution from Y.L. and B.J.E.; B.J.E. and Y.L. supervised the project.
\bibliography{main}

\begin{thebibliography}{22}%
\makeatletter
\providecommand \@ifxundefined [1]{%
 \@ifx{#1\undefined}
}%
\providecommand \@ifnum [1]{%
 \ifnum #1\expandafter \@firstoftwo
 \else \expandafter \@secondoftwo
 \fi
}%
\providecommand \@ifx [1]{%
 \ifx #1\expandafter \@firstoftwo
 \else \expandafter \@secondoftwo
 \fi
}%
\providecommand \natexlab [1]{#1}%
\providecommand \enquote  [1]{``#1''}%
\providecommand \bibnamefont  [1]{#1}%
\providecommand \bibfnamefont [1]{#1}%
\providecommand \citenamefont [1]{#1}%
\providecommand \href@noop [0]{\@secondoftwo}%
\providecommand \href [0]{\begingroup \@sanitize@url \@href}%
\providecommand \@href[1]{\@@startlink{#1}\@@href}%
\providecommand \@@href[1]{\endgroup#1\@@endlink}%
\providecommand \@sanitize@url [0]{\catcode `\\12\catcode `\$12\catcode
  `\&12\catcode `\#12\catcode `\^12\catcode `\_12\catcode `\%12\relax}%
\providecommand \@@startlink[1]{}%
\providecommand \@@endlink[0]{}%
\providecommand \url  [0]{\begingroup\@sanitize@url \@url }%
\providecommand \@url [1]{\endgroup\@href {#1}{\urlprefix }}%
\providecommand \urlprefix  [0]{URL }%
\providecommand \Eprint [0]{\href }%
\providecommand \doibase [0]{https://doi.org/}%
\providecommand \selectlanguage [0]{\@gobble}%
\providecommand \bibinfo  [0]{\@secondoftwo}%
\providecommand \bibfield  [0]{\@secondoftwo}%
\providecommand \translation [1]{[#1]}%
\providecommand \BibitemOpen [0]{}%
\providecommand \bibitemStop [0]{}%
\providecommand \bibitemNoStop [0]{.\EOS\space}%
\providecommand \EOS [0]{\spacefactor3000\relax}%
\providecommand \BibitemShut  [1]{\csname bibitem#1\endcsname}%
\let\auto@bib@innerbib\@empty
\bibitem [{\citenamefont {Duling}\ and\ \citenamefont
  {Zimdars}(2009)}]{Duling2009}%
  \BibitemOpen
  \bibfield  {author} {\bibinfo {author} {\bibfnamefont {I.}~\bibnamefont
  {Duling}}\ and\ \bibinfo {author} {\bibfnamefont {D.}~\bibnamefont
  {Zimdars}},\ }\bibfield  {title} {\bibinfo {title} {{Terahertz imaging:
  Revealing hidden defects}},\ }\href
  {https://doi.org/10.1038/nphoton.2009.206} {\bibfield  {journal} {\bibinfo
  {journal} {Nature Photonics}\ }\textbf {\bibinfo {volume} {3}},\ \bibinfo
  {pages} {630} (\bibinfo {year} {2009})}\BibitemShut {NoStop}%
\bibitem [{\citenamefont {Gowen}\ \emph {et~al.}(2012)\citenamefont {Gowen},
  \citenamefont {O'Sullivan},\ and\ \citenamefont {O'Donnell}}]{Gowen2012}%
  \BibitemOpen
  \bibfield  {author} {\bibinfo {author} {\bibfnamefont {A.~A.}\ \bibnamefont
  {Gowen}}, \bibinfo {author} {\bibfnamefont {C.}~\bibnamefont {O'Sullivan}},\
  and\ \bibinfo {author} {\bibfnamefont {C.~P.}\ \bibnamefont {O'Donnell}},\
  }\bibfield  {title} {\bibinfo {title} {{Terahertz time domain spectroscopy
  and imaging: Emerging techniques for food process monitoring and quality
  control}},\ }\href {https://doi.org/10.1016/j.tifs.2011.12.006} {\bibfield
  {journal} {\bibinfo  {journal} {Trends in Food Science and Technology}\
  }\textbf {\bibinfo {volume} {25}},\ \bibinfo {pages} {40} (\bibinfo {year}
  {2012})}\BibitemShut {NoStop}%
\bibitem [{\citenamefont {Caruso}\ \emph {et~al.}(2015)\citenamefont {Caruso},
  \citenamefont {Bassi}, \citenamefont {Bevilacqua},\ and\ \citenamefont
  {Neviani}}]{Caruso2015}%
  \BibitemOpen
  \bibfield  {author} {\bibinfo {author} {\bibfnamefont {M.}~\bibnamefont
  {Caruso}}, \bibinfo {author} {\bibfnamefont {M.}~\bibnamefont {Bassi}},
  \bibinfo {author} {\bibfnamefont {A.}~\bibnamefont {Bevilacqua}},\ and\
  \bibinfo {author} {\bibfnamefont {A.}~\bibnamefont {Neviani}},\ }\bibfield
  {title} {\bibinfo {title} {{A 2-16 GHz 65 nm CMOS stepped-frequency radar
  transmitter with harmonic rejection for high-resolution medical imaging
  applications}},\ }\href {https://doi.org/10.1109/TCSI.2014.2362332}
  {\bibfield  {journal} {\bibinfo  {journal} {IEEE Transactions on Circuits and
  Systems I: Regular Papers}\ }\textbf {\bibinfo {volume} {62}},\ \bibinfo
  {pages} {413} (\bibinfo {year} {2015})}\BibitemShut {NoStop}%
\bibitem [{\citenamefont {Richards}\ \emph {et~al.}(2010)\citenamefont
  {Richards}, \citenamefont {Scheer}, \citenamefont {Holm},\ and\ \citenamefont
  {Melvin}}]{Richards2010}%
  \BibitemOpen
  \bibfield  {author} {\bibinfo {author} {\bibfnamefont {M.~A.}\ \bibnamefont
  {Richards}}, \bibinfo {author} {\bibfnamefont {J.}~\bibnamefont {Scheer}},
  \bibinfo {author} {\bibfnamefont {W.~A.}\ \bibnamefont {Holm}},\ and\
  \bibinfo {author} {\bibfnamefont {W.~L.}\ \bibnamefont {Melvin}},\
  }\href@noop {} {\emph {\bibinfo {title} {{Principles of modern radar}}}}\
  (\bibinfo  {publisher} {Citeseer},\ \bibinfo {year} {2010})\BibitemShut
  {NoStop}%
\bibitem [{\citenamefont {Tonda-Goldstein}\ \emph {et~al.}(2006)\citenamefont
  {Tonda-Goldstein}, \citenamefont {Dolfi}, \citenamefont {Monsterleet},
  \citenamefont {Formont}, \citenamefont {Chazelas},\ and\ \citenamefont
  {Huignard}}]{Tonda-Goldstein2006}%
  \BibitemOpen
  \bibfield  {author} {\bibinfo {author} {\bibfnamefont {S.}~\bibnamefont
  {Tonda-Goldstein}}, \bibinfo {author} {\bibfnamefont {D.}~\bibnamefont
  {Dolfi}}, \bibinfo {author} {\bibfnamefont {A.}~\bibnamefont {Monsterleet}},
  \bibinfo {author} {\bibfnamefont {S.}~\bibnamefont {Formont}}, \bibinfo
  {author} {\bibfnamefont {J.}~\bibnamefont {Chazelas}},\ and\ \bibinfo
  {author} {\bibfnamefont {J.~P.}\ \bibnamefont {Huignard}},\ }\bibfield
  {title} {\bibinfo {title} {{Optical signal processing in radar systems}},\
  }\href {https://doi.org/10.1109/TMTT.2005.863059} {\bibfield  {journal}
  {\bibinfo  {journal} {IEEE Transactions on Microwave Theory and Techniques}\
  }\textbf {\bibinfo {volume} {54}},\ \bibinfo {pages} {847} (\bibinfo {year}
  {2006})}\BibitemShut {NoStop}%
\bibitem [{\citenamefont {Serafino}\ \emph {et~al.}(2020)\citenamefont
  {Serafino}, \citenamefont {Maresca}, \citenamefont {Porzi}, \citenamefont
  {Scotti}, \citenamefont {Ghelfi},\ and\ \citenamefont
  {Bogoni}}]{Serafino2020a}%
  \BibitemOpen
  \bibfield  {author} {\bibinfo {author} {\bibfnamefont {G.}~\bibnamefont
  {Serafino}}, \bibinfo {author} {\bibfnamefont {S.}~\bibnamefont {Maresca}},
  \bibinfo {author} {\bibfnamefont {C.}~\bibnamefont {Porzi}}, \bibinfo
  {author} {\bibfnamefont {F.}~\bibnamefont {Scotti}}, \bibinfo {author}
  {\bibfnamefont {P.}~\bibnamefont {Ghelfi}},\ and\ \bibinfo {author}
  {\bibfnamefont {A.}~\bibnamefont {Bogoni}},\ }\bibfield  {title} {\bibinfo
  {title} {{Microwave Photonics for Remote Sensing: From Basic Concepts to
  High-Level Functionalities}},\ }\href
  {https://doi.org/10.1109/JLT.2020.2989618} {\bibfield  {journal} {\bibinfo
  {journal} {Journal of Lightwave Technology}\ }\textbf {\bibinfo {volume}
  {38}},\ \bibinfo {pages} {5339} (\bibinfo {year} {2020})}\BibitemShut
  {NoStop}%
\bibitem [{\citenamefont {Pan}\ and\ \citenamefont {Zhang}(2020)}]{Pan2020}%
  \BibitemOpen
  \bibfield  {author} {\bibinfo {author} {\bibfnamefont {S.}~\bibnamefont
  {Pan}}\ and\ \bibinfo {author} {\bibfnamefont {Y.}~\bibnamefont {Zhang}},\
  }\bibfield  {title} {\bibinfo {title} {{Microwave Photonic Radars}},\ }\href
  {https://doi.org/10.1109/JLT.2020.2993166} {\bibfield  {journal} {\bibinfo
  {journal} {Journal of Lightwave Technology}\ }\textbf {\bibinfo {volume}
  {38}},\ \bibinfo {pages} {5450} (\bibinfo {year} {2020})}\BibitemShut
  {NoStop}%
\bibitem [{\citenamefont {Kittlaus}\ \emph {et~al.}(2021)\citenamefont
  {Kittlaus}, \citenamefont {Eliyahu}, \citenamefont {Ganji}, \citenamefont
  {Williams}, \citenamefont {Matsko}, \citenamefont {Cooper},\ and\
  \citenamefont {Forouhar}}]{Kittlaus2021}%
  \BibitemOpen
  \bibfield  {author} {\bibinfo {author} {\bibfnamefont {E.~A.}\ \bibnamefont
  {Kittlaus}}, \bibinfo {author} {\bibfnamefont {D.}~\bibnamefont {Eliyahu}},
  \bibinfo {author} {\bibfnamefont {S.}~\bibnamefont {Ganji}}, \bibinfo
  {author} {\bibfnamefont {S.}~\bibnamefont {Williams}}, \bibinfo {author}
  {\bibfnamefont {A.~B.}\ \bibnamefont {Matsko}}, \bibinfo {author}
  {\bibfnamefont {K.~B.}\ \bibnamefont {Cooper}},\ and\ \bibinfo {author}
  {\bibfnamefont {S.}~\bibnamefont {Forouhar}},\ }\bibfield  {title} {\bibinfo
  {title} {{A low-noise photonic heterodyne synthesizer and its application to
  millimeter-wave radar}},\ }\href {https://doi.org/10.1038/s41467-021-24637-0}
  {\bibfield  {journal} {\bibinfo  {journal} {Nature Communications}\ }\textbf
  {\bibinfo {volume} {12}},\ \bibinfo {pages} {1} (\bibinfo {year}
  {2021})}\BibitemShut {NoStop}%
\bibitem [{\citenamefont {Serafino}\ \emph {et~al.}(2019)\citenamefont
  {Serafino}, \citenamefont {Scotti}, \citenamefont {Lembo}, \citenamefont
  {Hussain}, \citenamefont {Porzi}, \citenamefont {Malacarne}, \citenamefont
  {Maresca}, \citenamefont {Onori}, \citenamefont {Ghelfi},\ and\ \citenamefont
  {Bogoni}}]{Serafino2019}%
  \BibitemOpen
  \bibfield  {author} {\bibinfo {author} {\bibfnamefont {G.}~\bibnamefont
  {Serafino}}, \bibinfo {author} {\bibfnamefont {F.}~\bibnamefont {Scotti}},
  \bibinfo {author} {\bibfnamefont {L.}~\bibnamefont {Lembo}}, \bibinfo
  {author} {\bibfnamefont {B.}~\bibnamefont {Hussain}}, \bibinfo {author}
  {\bibfnamefont {C.}~\bibnamefont {Porzi}}, \bibinfo {author} {\bibfnamefont
  {A.}~\bibnamefont {Malacarne}}, \bibinfo {author} {\bibfnamefont
  {S.}~\bibnamefont {Maresca}}, \bibinfo {author} {\bibfnamefont
  {D.}~\bibnamefont {Onori}}, \bibinfo {author} {\bibfnamefont
  {P.}~\bibnamefont {Ghelfi}},\ and\ \bibinfo {author} {\bibfnamefont
  {A.}~\bibnamefont {Bogoni}},\ }\bibfield  {title} {\bibinfo {title} {{Toward
  a new generation of radar systems based on microwave photonic
  technologies}},\ }\href {https://doi.org/10.1109/JLT.2019.2894224} {\bibfield
   {journal} {\bibinfo  {journal} {Journal of Lightwave Technology}\ }\textbf
  {\bibinfo {volume} {37}},\ \bibinfo {pages} {643} (\bibinfo {year}
  {2019})}\BibitemShut {NoStop}%
\bibitem [{\citenamefont {Fu}\ \emph {et~al.}(2013)\citenamefont {Fu},
  \citenamefont {Zhang}, \citenamefont {Hraimel}, \citenamefont {Liu},\ and\
  \citenamefont {Shen}}]{Fu2013}%
  \BibitemOpen
  \bibfield  {author} {\bibinfo {author} {\bibfnamefont {Y.}~\bibnamefont
  {Fu}}, \bibinfo {author} {\bibfnamefont {X.}~\bibnamefont {Zhang}}, \bibinfo
  {author} {\bibfnamefont {B.}~\bibnamefont {Hraimel}}, \bibinfo {author}
  {\bibfnamefont {T.}~\bibnamefont {Liu}},\ and\ \bibinfo {author}
  {\bibfnamefont {D.}~\bibnamefont {Shen}},\ }\bibfield  {title} {\bibinfo
  {title} {{Mach-Zehnder: A review of bias control techniques for mach-zehnder
  modulators in photonic analog links}},\ }\href
  {https://doi.org/10.1109/MMM.2013.2280332} {\bibfield  {journal} {\bibinfo
  {journal} {IEEE Microwave Magazine}\ }\textbf {\bibinfo {volume} {14}},\
  \bibinfo {pages} {102} (\bibinfo {year} {2013})}\BibitemShut {NoStop}%
\bibitem [{\citenamefont {Zhang}\ and\ \citenamefont {Yao}(2014)}]{Zhang2014}%
  \BibitemOpen
  \bibfield  {author} {\bibinfo {author} {\bibfnamefont {J.}~\bibnamefont
  {Zhang}}\ and\ \bibinfo {author} {\bibfnamefont {J.}~\bibnamefont {Yao}},\
  }\bibfield  {title} {\bibinfo {title} {{Time-stretched sampling of a fast
  microwave waveform based on the repetitive use of a linearly chirped fiber
  Bragg grating in a dispersive loop}},\ }\href
  {https://doi.org/10.1364/optica.1.000064} {\bibfield  {journal} {\bibinfo
  {journal} {Optica}\ }\textbf {\bibinfo {volume} {1}},\ \bibinfo {pages} {64}
  (\bibinfo {year} {2014})}\BibitemShut {NoStop}%
\bibitem [{\citenamefont {Zhou}\ \emph {et~al.}(2016)\citenamefont {Zhou},
  \citenamefont {Zhang}, \citenamefont {Guo},\ and\ \citenamefont
  {Pan}}]{Zhou2016}%
  \BibitemOpen
  \bibfield  {author} {\bibinfo {author} {\bibfnamefont {P.}~\bibnamefont
  {Zhou}}, \bibinfo {author} {\bibfnamefont {F.}~\bibnamefont {Zhang}},
  \bibinfo {author} {\bibfnamefont {Q.}~\bibnamefont {Guo}},\ and\ \bibinfo
  {author} {\bibfnamefont {S.}~\bibnamefont {Pan}},\ }\bibfield  {title}
  {\bibinfo {title} {{Linearly chirped microwave waveform generation with large
  time-bandwidth product by optically injected semiconductor laser}},\ }\href
  {https://doi.org/10.1364/oe.24.018460} {\bibfield  {journal} {\bibinfo
  {journal} {Optics Express}\ }\textbf {\bibinfo {volume} {24}},\ \bibinfo
  {pages} {18460} (\bibinfo {year} {2016})}\BibitemShut {NoStop}%
\bibitem [{\citenamefont {Hao}\ \emph {et~al.}(2018)\citenamefont {Hao},
  \citenamefont {Cen}, \citenamefont {Dai}, \citenamefont {Tang}, \citenamefont
  {Li}, \citenamefont {Yao}, \citenamefont {Zhu},\ and\ \citenamefont
  {Li}}]{Hao2018}%
  \BibitemOpen
  \bibfield  {author} {\bibinfo {author} {\bibfnamefont {T.}~\bibnamefont
  {Hao}}, \bibinfo {author} {\bibfnamefont {Q.}~\bibnamefont {Cen}}, \bibinfo
  {author} {\bibfnamefont {Y.}~\bibnamefont {Dai}}, \bibinfo {author}
  {\bibfnamefont {J.}~\bibnamefont {Tang}}, \bibinfo {author} {\bibfnamefont
  {W.}~\bibnamefont {Li}}, \bibinfo {author} {\bibfnamefont {J.}~\bibnamefont
  {Yao}}, \bibinfo {author} {\bibfnamefont {N.}~\bibnamefont {Zhu}},\ and\
  \bibinfo {author} {\bibfnamefont {M.}~\bibnamefont {Li}},\ }\bibfield
  {title} {\bibinfo {title} {{Breaking the limitation of mode building time in
  an optoelectronic oscillator}},\ }\bibfield  {journal} {\bibinfo  {journal}
  {Nature Communications}\ }\textbf {\bibinfo {volume} {9}},\ \href
  {https://doi.org/10.1038/s41467-018-04240-6} {10.1038/s41467-018-04240-6}
  (\bibinfo {year} {2018})\BibitemShut {NoStop}%
\bibitem [{\citenamefont {Zhang}\ \emph {et~al.}(2020)\citenamefont {Zhang},
  \citenamefont {Liu}, \citenamefont {Burla},\ and\ \citenamefont
  {Eggleton}}]{Zhang2020a}%
  \BibitemOpen
  \bibfield  {author} {\bibinfo {author} {\bibfnamefont {Z.}~\bibnamefont
  {Zhang}}, \bibinfo {author} {\bibfnamefont {Y.}~\bibnamefont {Liu}}, \bibinfo
  {author} {\bibfnamefont {M.}~\bibnamefont {Burla}},\ and\ \bibinfo {author}
  {\bibfnamefont {B.~J.}\ \bibnamefont {Eggleton}},\ }\bibfield  {title}
  {\bibinfo {title} {{5.6-GHz-Bandwidth Photonic Stepped-Frequency Radar using
  MHz-Level Frequency-Shifting Modulation}},\ }in\ \href@noop {} {\emph
  {\bibinfo {booktitle} {2020 Conference on Lasers and Electro-Optics
  (CLEO)}}}\ (\bibinfo {year} {2020})\ pp.\ \bibinfo {pages} {1--2}\BibitemShut
  {NoStop}%
\bibitem [{\citenamefont {Zhang}\ \emph {et~al.}(2021)\citenamefont {Zhang},
  \citenamefont {Liu}, \citenamefont {Zhang}, \citenamefont {Shao},
  \citenamefont {Ma}, \citenamefont {Li}, \citenamefont {Sun}, \citenamefont
  {Li},\ and\ \citenamefont {Pan}}]{Zhang2021}%
  \BibitemOpen
  \bibfield  {author} {\bibinfo {author} {\bibfnamefont {Y.}~\bibnamefont
  {Zhang}}, \bibinfo {author} {\bibfnamefont {C.}~\bibnamefont {Liu}}, \bibinfo
  {author} {\bibfnamefont {Y.}~\bibnamefont {Zhang}}, \bibinfo {author}
  {\bibfnamefont {K.}~\bibnamefont {Shao}}, \bibinfo {author} {\bibfnamefont
  {C.}~\bibnamefont {Ma}}, \bibinfo {author} {\bibfnamefont {L.}~\bibnamefont
  {Li}}, \bibinfo {author} {\bibfnamefont {L.}~\bibnamefont {Sun}}, \bibinfo
  {author} {\bibfnamefont {S.}~\bibnamefont {Li}},\ and\ \bibinfo {author}
  {\bibfnamefont {S.}~\bibnamefont {Pan}},\ }\bibfield  {title} {\bibinfo
  {title} {{Multi-Functional Radar Waveform Generation Based on Optical
  Frequency-Time Stitching Method}},\ }\href
  {https://doi.org/10.1109/JLT.2020.3029275} {\bibfield  {journal} {\bibinfo
  {journal} {Journal of Lightwave Technology}\ }\textbf {\bibinfo {volume}
  {39}},\ \bibinfo {pages} {458} (\bibinfo {year} {2021})}\BibitemShut
  {NoStop}%
\bibitem [{\citenamefont {Kikuchi}(2016)}]{Kikuchi2016}%
  \BibitemOpen
  \bibfield  {author} {\bibinfo {author} {\bibfnamefont {K.}~\bibnamefont
  {Kikuchi}},\ }\bibfield  {title} {\bibinfo {title} {{Fundamentals of coherent
  optical fiber communications}},\ }\href
  {https://doi.org/10.1109/JLT.2015.2463719} {\bibfield  {journal} {\bibinfo
  {journal} {Journal of Lightwave Technology}\ }\textbf {\bibinfo {volume}
  {34}},\ \bibinfo {pages} {157} (\bibinfo {year} {2016})}\BibitemShut
  {NoStop}%
\bibitem [{\citenamefont {Ozdemir}(2012)}]{Ozdemir2012}%
  \BibitemOpen
  \bibfield  {author} {\bibinfo {author} {\bibfnamefont {C.}~\bibnamefont
  {Ozdemir}},\ }\href
  {http://ebookcentral.proquest.com/lib/usyd/detail.action?docID=818515} {\emph
  {\bibinfo {title} {{Inverse Synthetic Aperture Radar Imaging with MATLAB
  Algorithms}}}}\ (\bibinfo  {publisher} {John Wiley {\&} Sons, Incorporated},\
  \bibinfo {address} {Hoboken, UNITED STATES},\ \bibinfo {year} {2012})\ pp.\
  \bibinfo {pages} {57--60}\BibitemShut {NoStop}%
\bibitem [{\citenamefont {Levanon}(2002)}]{Levanon2002}%
  \BibitemOpen
  \bibfield  {author} {\bibinfo {author} {\bibfnamefont {N.}~\bibnamefont
  {Levanon}},\ }\bibfield  {title} {\bibinfo {title} {{Stepped-frequency
  pulse-train radar signal}},\ }\href {https://doi.org/10.1049/ip-rsn:20020432}
  {\bibfield  {journal} {\bibinfo  {journal} {IEE Proceedings: Radar, Sonar and
  Navigation}\ ,\ \bibinfo {pages} {297}} (\bibinfo {year} {2002})}\BibitemShut
  {NoStop}%
\bibitem [{\citenamefont {Jia}\ \emph {et~al.}(2021)\citenamefont {Jia},
  \citenamefont {Guo}, \citenamefont {Wang}, \citenamefont {Song},
  \citenamefont {Cui},\ and\ \citenamefont {Zhong}}]{Jia2021a}%
  \BibitemOpen
  \bibfield  {author} {\bibinfo {author} {\bibfnamefont {Y.}~\bibnamefont
  {Jia}}, \bibinfo {author} {\bibfnamefont {Y.}~\bibnamefont {Guo}}, \bibinfo
  {author} {\bibfnamefont {G.}~\bibnamefont {Wang}}, \bibinfo {author}
  {\bibfnamefont {R.}~\bibnamefont {Song}}, \bibinfo {author} {\bibfnamefont
  {G.}~\bibnamefont {Cui}},\ and\ \bibinfo {author} {\bibfnamefont
  {X.}~\bibnamefont {Zhong}},\ }\bibfield  {title} {\bibinfo {title}
  {{Multi-frequency and multi-domain human activity recognition based on SFCW
  radar using deep learning}},\ }\href
  {https://doi.org/https://doi.org/10.1016/j.neucom.2020.07.136} {\bibfield
  {journal} {\bibinfo  {journal} {Neurocomputing}\ }\textbf {\bibinfo {volume}
  {444}},\ \bibinfo {pages} {274} (\bibinfo {year} {2021})}\BibitemShut
  {NoStop}%
\bibitem [{\citenamefont {Billault}\ \emph {et~al.}(2021)\citenamefont
  {Billault}, \citenamefont {Arpison}, \citenamefont {Crozatier}, \citenamefont
  {Kemlin}, \citenamefont {Morvan}, \citenamefont {Dolfi},\ and\ \citenamefont
  {De~Chatellus}}]{Billault2021}%
  \BibitemOpen
  \bibfield  {author} {\bibinfo {author} {\bibfnamefont {V.}~\bibnamefont
  {Billault}}, \bibinfo {author} {\bibfnamefont {G.}~\bibnamefont {Arpison}},
  \bibinfo {author} {\bibfnamefont {V.}~\bibnamefont {Crozatier}}, \bibinfo
  {author} {\bibfnamefont {V.}~\bibnamefont {Kemlin}}, \bibinfo {author}
  {\bibfnamefont {L.}~\bibnamefont {Morvan}}, \bibinfo {author} {\bibfnamefont
  {D.}~\bibnamefont {Dolfi}},\ and\ \bibinfo {author} {\bibfnamefont {H.~G.}\
  \bibnamefont {De~Chatellus}},\ }\bibfield  {title} {\bibinfo {title}
  {{Coherent Optical Fiber Sensing Based on a Frequency Shifting Loop}},\
  }\href {https://doi.org/10.1109/JLT.2021.3060105} {\bibfield  {journal}
  {\bibinfo  {journal} {Journal of Lightwave Technology}\ }\textbf {\bibinfo
  {volume} {39}},\ \bibinfo {pages} {4118} (\bibinfo {year}
  {2021})}\BibitemShut {NoStop}%
\bibitem [{\citenamefont {Gustavsson}\ \emph {et~al.}(1990)\citenamefont
  {Gustavsson}, \citenamefont {Frklind}, \citenamefont {Hellsten},
  \citenamefont {Jonsson}, \citenamefont {Larsson},\ and\ \citenamefont
  {Stenstrom}}]{Gustavsson1990}%
  \BibitemOpen
  \bibfield  {author} {\bibinfo {author} {\bibfnamefont {A.}~\bibnamefont
  {Gustavsson}}, \bibinfo {author} {\bibfnamefont {P.}~\bibnamefont {Frklind}},
  \bibinfo {author} {\bibfnamefont {H.}~\bibnamefont {Hellsten}}, \bibinfo
  {author} {\bibfnamefont {T.}~\bibnamefont {Jonsson}}, \bibinfo {author}
  {\bibfnamefont {B.}~\bibnamefont {Larsson}},\ and\ \bibinfo {author}
  {\bibfnamefont {G.}~\bibnamefont {Stenstrom}},\ }\bibfield  {title} {\bibinfo
  {title} {{The Airborne VHF SAR System CARABAS}},\ }\href@noop {} {\bibfield
  {journal} {\bibinfo  {journal} {IEEE International Geoscience and Remote
  Sensing Symposium}\ ,\ \bibinfo {pages} {558}} (\bibinfo {year}
  {1990})}\BibitemShut {NoStop}%
\bibitem [{\citenamefont {Shao}\ \emph {et~al.}(2020)\citenamefont {Shao},
  \citenamefont {Sinclair}, \citenamefont {Leatham}, \citenamefont {Hu},
  \citenamefont {Yu}, \citenamefont {Turpin}, \citenamefont {Crowe},\ and\
  \citenamefont {Lon{\v{c}}ar}}]{Shao2020a}%
  \BibitemOpen
  \bibfield  {author} {\bibinfo {author} {\bibfnamefont {L.}~\bibnamefont
  {Shao}}, \bibinfo {author} {\bibfnamefont {N.}~\bibnamefont {Sinclair}},
  \bibinfo {author} {\bibfnamefont {J.}~\bibnamefont {Leatham}}, \bibinfo
  {author} {\bibfnamefont {Y.}~\bibnamefont {Hu}}, \bibinfo {author}
  {\bibfnamefont {M.}~\bibnamefont {Yu}}, \bibinfo {author} {\bibfnamefont
  {T.}~\bibnamefont {Turpin}}, \bibinfo {author} {\bibfnamefont
  {D.}~\bibnamefont {Crowe}},\ and\ \bibinfo {author} {\bibfnamefont
  {M.}~\bibnamefont {Lon{\v{c}}ar}},\ }\bibfield  {title} {\bibinfo {title}
  {{Integrated microwave acousto-optic frequency shifter on thin-film lithium
  niobate}},\ }\href {https://doi.org/10.1364/oe.397138} {\bibfield  {journal}
  {\bibinfo  {journal} {Optics Express}\ }\textbf {\bibinfo {volume} {28}},\
  \bibinfo {pages} {23728} (\bibinfo {year} {2020})}\BibitemShut {NoStop}%
\end{thebibliography}%

\end{document}